\documentclass[11pt, oneside]{article}   	
\usepackage{geometry}                		
\usepackage[parfill]{parskip}    		
\usepackage{graphicx}				
\usepackage{authblk}
\usepackage{amssymb}
\usepackage{amsmath}
\usepackage{amsfonts}
\usepackage{mathbbol}

\title{Embedding Cosmology and Gravity}
\author{Abhishek Goswami\footnote{
This idea was developed while I was a graduate student in the physics and mathematics departments at 
SUNY Buffalo. The work was concluded while I was working as an adjunct instructor in the mathematics
department at SUNY Buffalo after my graduation. 
\newline \textbf{e-mail}: goswami3@buffalo.edu }}
\date{}							

\begin{document}
\maketitle
\begin{abstract}
I start with a scenario where the universe is an abstract space $\mathcal{M}$ having \textit{d} dimensions. 
 There is a two dimensional surface embedded in it. Embedding is a map from the embedded surface to
 $\mathcal{M}$ that has a field theory described by Sigma model.
 I take \textit{d} directions of $\mathcal{M}$ to be the generators of a symmetry group SU(n) of
 the Lagrangian of the embedding. This means embedding has n flavors.
 Then I introduce spontaneous symmetry breaking in the theory and define the direction along which the
 symmetry breaking occurs as time. Next I write down the modified Einstein's equation including the embedding. 
 Then I discuss embedding's relation to the expansion of the universe. After that I construct an inflationary scenario with 
 embedding as inflaton and discuss its connection to Starobinsky $R^{2}$ model. Finally, I discuss the effect of inflation 
 on the non-commutativity of the spacetime.
\end{abstract}
 
\tableofcontents

\section{Introduction}
\label{intro}
Cosmology at present is studied using the Hot Big Bang model that is constructed from Einstein's general relativity.
It rests firmly on three pillars: (i) Big Bang Nucleosynthesis,
(ii) Cosmic Microwave Background Radiation and (iii) Hubble Redshift. At Planck scale ($t \sim 10^{-43} s$),
we have a hot ($T \sim 10^{32}$ K) and dense universe
which is expanding and cooling. All the regions are equivalent and there is no special point or center of the universe. 
Therefore, the Big Bang must have occurred everywhere at once. 

The current state of the art in cosmology
has two approaches to study the universe at Planck scale. The first is an inflationary paradigm 
\cite{Star, G, L1, AS, L2} which is based on
two assumptions: (i) a potential landscape of some field (ii) with the field sitting in some metastable vacuum state.
Hence, in this approach the hot universe is taken to be filled with some vacuum like energy density which is in
a metastable state (false vacuum). Then the phase transition from this false vacuum (at $ t \sim 10^{-36} s$) in a 
homogeneous patch can trigger inflation and the subsequent multiverse.
The origin of this potential landscape is unknown and is attributed to the lack of proper initial conditions for
an inflationary scenario. Therefore, the inflationary picture, despite its successes (see for instance, \cite{L3})
is widely considered to be an incomplete one.

The second approach explores the quantum origin of the universe and tries to explain the Big Bang itself.
For example, a universe created by quantum tunneling from nothing \cite{V, Rub, ZS} precedes an inflationary scenario.
Another one is a cyclic universe model that replaces Big Bang with a Big Bounce \cite{ST} and provides an
alternative to the inflation paradigm. What happened at the Big Bang (or the physics before $t \sim 10^{-43} s$)
remains an open problem.

Another important aspect of the Hot Big Bang model is the presence of a dark sector. As per the observations
\cite{Planck} the universe has $\sim 68\%$ of dark energy and $\sim 27\%$ of dark matter content. Dark energy is 
most commonly studied by adding a cosmological constant $\Lambda$ in Einstein's equations as
vacuum energy density. However, its origin is still unknown. Moreover while trying to calculate its observed 
value using vacuum energy of quantum field theory we are led to what is known as \lq\lq vacuum catastrophe\rq\rq \cite{ACJ}.
Dark matter models involve both a particle physics side and a modified gravity approach. 
However, there is no experimental evidence yet of dark matter candidates such as WIMPs \cite{WIMP}.
Modified Newtonian Dynamics (MOND) cannot fully eliminate the need for dark matter \cite{MOND}. 
Also MOND's prediction that gravitational waves should travel at a speed different than the speed of light
is now ruled out \cite{BDKW}. Thus, dark matter is still a mystery.

The cosmological arrow of time is central to the Hot Big Bang model. It is 
addressed using entropy considerations, see for example, \cite{DKS}. 
As per the second law of thermodynamics the entropy of a closed system almost always increases. 
Since the laws of physics allow processes with a reverse arrow of time, the second law is then understood 
to give a preferred direction to the events in nature. The direction being the one that increases entropy of the universe.
However, this just translates the problem of arrow of time to a different problem; why did the 
universe start in a very low entropy state? This is an open problem.

Here, I address the Big Bang singularity by noting that the singularity is purely temporal in nature due to
an arrow of time. Thus, the universe before the Big Bang should be understood as an abstract space.
The arrow of time (meaning time's asymmetry) is a physical fact of nature and emerges from this abstract space via a Big Bang.
Thus, Big Bang makes one of the axes of the abstract space asymmetrical giving a geometric explanation of arrow of time.
Note that due to the uncertainty relation $\Delta E \Delta t \geqslant \hbar$, a non zero energy would accompany
the emergence of time. By doing so I also seek to advance our understanding of the dark sector.

Before the Big Bang assume that the universe is an abstract space $\mathcal{M}$ 
having $d$ dimensions and no time dimension; $(0, d)$. There is a
two dimensional surface (dark surface) embedded in it. Embedding is a map from the dark surface
parametrized by $(u,v)$ to the target space $\mathcal{M}$ and can be treated as a scalar field; $\phi$
(for example, Sigma model). 
\begin{equation}
\phi^{i} : (u,v) \rightarrow X^{\mu, i}(u,v)
\end{equation}
where $i = 1,2, \cdots, \text{n}$ denotes some internal symmetry of embeddings and $X^{\mu}$ are space
coordinates of $\mathcal{M}$ with $\mu = 1,2, \cdots, d$. Denote $\Phi = (\phi_{1},\cdots, \phi_{\text{n}})^{\dag}$.
The Lagrangian of the embedding is given by
\begin{equation}
\label{2}
\mathcal{L} = \frac{1}{2}(\partial_{a}\Phi)^{\dag}(\partial^{a}\Phi) 
+ \frac{\mu^{2}}{2} \Phi^{\dag}\Phi - \frac{\lambda}{4}(\Phi^{\dag}\Phi)^{2}
\end{equation}
where $a$ denotes the coordinates of the dark surface; $(u,v)$. For $\Phi \rightarrow \mathrm{G} \Phi$, 
where $\mathrm{G} \in$ SU(n), $\mathcal{L}$ is invariant. Then
$X^{\mu}$ are the generators of a symmetry group SU(n) of the Lagrangian theory of embedding Eq. (\ref{2}) such that 
$d = \text{n}^{2} - 1$. Let $\Phi$ acquires a VEV
\begin{equation}
\label{3}
\langle \Phi \rangle = \frac{\mu}{\sqrt{\lambda}} \hspace{0.05 cm} X^{t}.
\end{equation}
The SU(n) symmetry is then spontaneously broken. Define the direction along which symmetry breaking
occurs; $X^{t}$ as time. The closed (unitary) subspace of coordinate axes; $X^{\mu}$ satisfying
\begin{equation}
\label{4}
[X^{t}, X^{\mu}] = 0
\end{equation}
form classical spacetime. While any two coordinates $X^{\mu}$ and $X^{\nu}$ satisfying Eq. (\ref{4}) also satisfy
\begin{equation}
\label{5}
[X^{\mu}, X^{\nu}] = i \sum_{\alpha} f^{\alpha\mu\nu} X^{\alpha}
\end{equation}
where $f^{\alpha\mu\nu}$ are known as structure constants. Eq. (\ref{4}) and Eq. (\ref{5}) together is the non-commutative spacetime
with $X^{t}$ being the arrow of time. Embedding acquiring a VEV is what corresponds to the Big Bang while
the embedded surface belongs to the dark sector.

To better understand how the symmetry breaking Eq. (\ref{3}) changes the signature of the bulk space $\mathcal{M}$
consider $n = 2$ case. The Lagrangian Eq. (\ref{2}) has a SU(2) symmetry such that for any $\mathrm{G} \in$ SU(2),
$\Phi \rightarrow \mathrm{G} \Phi$ keeps $\mathcal{L}$ invariant. Then the bulk space $\mathcal{M}$ is a \textit{3-sphere}.
Let $T^{1}, T^{2}, T^{3}$ denote the generators of SU(2). These generators transform (or rotate) the vacuum state of embedding 
(that takes values on \textit{3-sphere} $\mathcal{M}$) along the specified axis. When $\langle\Phi\rangle = 0$, the vacuum state of 
embedding (hence the $\mathcal{M}$) remains invariant under the action of $T^{1}, T^{2}, T^{3}$. This is SU(2) symmetry as all the 
3 directions are same for the vacuum state. However, if $\langle\Phi\rangle = \pm \frac{\mu}{\sqrt{\lambda}} \hspace{0.05 cm} T^{3}$ 
then only the action of generator $T^{3}$ (rotation about the axis 3) keeps the vacuum state invariant.
This is because $[T^{1}, T^{3}] \neq 0$ and $[T^{2}, T^{3}] \neq 0$. This means the bulk space $\mathcal{M}$ 
now only has a U(1) symmetry. Note that rotation about $T^{3}$ means that $T^{3}$ remains fixed.
Thus, any transformation that leaves the vacuum state invariant also keeps $T^{3}$ fixed. 
Selecting $\langle\Phi\rangle = + \frac{\mu}{\sqrt{\lambda}} \hspace{0.05 cm} T^{3}$, also fixes the orientation of
$T^{3}$ axis. Thus, due to these two simultaneous constraints
$T^{3}$ now acts like a time axis in $\mathcal{M}$ that has a fixed alignment (hence the arrow). 
This changes the signature of $\mathcal{M}$. Note that for a general case of SU(n), rotation about the axes
satisfying Eq. (\ref{4}) also keeps the vacuum state (hence $\mathcal{M}$) invariant but the axes are spacelike. 
Only axis $X^{t}$ of Eq. (\ref{3}) has a fixed orientation (hence the arrow) as well and therefore, is timelike.

Embedding cosmology discussed here is not same as \textit{braneworld} cosmology and thus, should not be confused with it.
In braneworld cosmology models, our visible universe lives on a $(1+ 3)$ dimensional surface known as \textit{brane} embedded in
a higher dimensional spacetime known as the \textit{bulk}. Such models with $(1 + 3 + N)$ dimensions of the bulk, where
$N$ is the number of extra spacelike dimensions, have been extensively studied in cosmology,
see for instance, Randall-Sundrum models involving five dimensional bulk spacetime \cite{RS1, RS2}.
In braneworld models, the \textit{embedding} of \textit{brane} in the \textit{bulk} spacetime is characterized by the extrinsic curvature
of the embedded surface i.e. brane. Embedding as a map is not being studied in the braneworld cosmology. 
Here, I treat embedding as a map as described by the Sigma model framework instead of characterizing it by
extrinsic curvature of the embedded surface. Embedding map is a field that is defined on the embedded surface
inside the bulk. This embedded surface is not a brane on which our visible universe is confined. 
Instead it is some new abstract surface which is part of the 
dark sector and is embedded in the bulk. Studying embedding as a map can provide a physical explanation 
for the important open problems in cosmology. This is what is being attempted here independent of the 
braneworld models. Note that after the symmetry breaking Eq. (\ref{3}) the bulk space $\mathcal{M}$
becomes the bulk spacetime $\mathcal{M}$ and can be studied using the existing brane cosmology framework
independently. The key difference between the embedding and the braneworld cosmologies is how the embedding is 
characterized.

In section \ref{sigma}, I give a brief overview of the Sigma model framework in physics. Section \ref{embedding} provides more details of
embedding as new physics by deriving modified Einstein's equations including embedding. Section \ref{discussion}
includes some discussion on embedding's connection to the expansion of the universe, an inflationary scenario with
embedding as inflaton and effect of inflation on the non-commutativity of the spacetime. Finally, I conclude my
work in section \ref{conclusion}. 

\section{Sigma model framework}
\label{sigma}
Sigma model is the study of embedding of a surface $\Sigma$ (base manifold) in a target space $\mathcal{T}$.
It was first introduced by Gell-mann and L$\acute{e}$vy \cite{GL} while studying the beta decay to describe a particle
$\sigma$ that took values in some manifold. In the first versions of the sigma model $\Sigma$ is taken
to be the spacetime with coordinates $x^{\mu}$, where $\mu = 1, \cdots, d$ i.e. $d$ dimensional base manifold
and embedding is defined as 
\begin{equation}
\phi^{i} : x^{\mu} \rightarrow \phi^{i}(x^{\mu})
\end{equation}
where $\phi^{i}$ are the coordinates of $\mathcal{T}$ with $i = 1, \cdots, n$ i.e. $n$ dimensional target space. 
The action is given by
\begin{equation}
S = \int d^{d}x \Big[ \frac{1}{2} g_{ij}(\phi) \partial_{\mu} \phi^{i} \partial^{\mu} \phi^{j} + V(\phi) \Big]
\end{equation}
where $g_{ij}$ is a Reimannian metric on $\mathcal{T}$. It is common to take the target space as some
Lie group for example, O(n).

In string theory framework, $\Sigma$ is taken to be the string worldsheet i.e. trajectory traced out by a string
and $\mathcal{T}$ is taken to be the spacetime. Embedding is denoted as $X^{\mu} (\sigma, \tau)$, where
$\sigma, \tau$ are the local coordinates of the worldsheet and $X^{\mu}$ are spacetime coordinates.
The Polyakov action is given by
\begin{equation}
S_{P} = \frac{T}{2} \int d\sigma d\tau \sqrt{\gamma}
 \gamma^{ab} \partial_{a} X^{\mu} \partial_{b} X^{\nu} G_{\mu\nu} (X)
\end{equation}
where $\gamma_{ab}$ is the worldsheet metric, $G_{\mu\nu} (X)$ is the spacetime metric and $T$ is the
string tension. For more applications of sigma model see for example, WZNW model \cite{WZ, N, W}.

Here, I apply Sigma model framework in a completely new manner giving a new physical meaning to embedding.
The details follow in the next section.

\section{Embedding as new physics}
\label{embedding}
For a better understanding of the new physics due to embedding first consider a simple analogy. A function $f$ is defined
from a domain set $\{x\}$ to a range set \{y\} as
\begin{center}
$f : x \rightarrow y$
\end{center}
such that $y = f(x)$. Then it is common to use $y$ and $f(x)$ interchangeably as they are taken to represent the 
same thing. Just like in string theory embedding is identified as $X^{\mu}(\sigma, \tau)$. While there is nothing
wrong with this identification but it hides a deeper physics beneath it. Here, we explore that by giving a physical
meaning to $f$ itself instead of just identifying it with $y$.

$\mathcal{M}$ is a $(0, d)$ manifold which is our target space (and will eventually be the spacetime).
$\Phi$ is the embedding of a two dimensional surface $(u, v)$ (which will be a part of the dark sector) in $\mathcal{M}$.
Let 
\begin{center}
$\Phi = (\phi_{1}, \cdots, \phi_{n})^{\dag}$ \hspace{0.5 cm} $\phi^{i} : (u, v) \rightarrow X^{\mu, i}(u, v)$
\end{center}
where $\mu = 1, \cdots d$, $i=1, \cdots, n$ and each $\phi^{i}$ is a doublet\footnote{A more precise notation
would be $\phi_{i_{k}} = \phi_{i_{1}} + i\hspace{0.05 cm} \phi_{i_{2}}$ with  $ \phi^{i_{k}} : (u, v) \rightarrow X^{\mu, i_{k}}(u, v)$.
For the purpose of clarity we do not use another index $k$ throughout the paper. We assume that the reader would
regard $\phi$ as a doublet only.}.
Note that the coordinates of $\mathcal{M}$, $X$ have two indices; $\mu$ and $i$. Denote

\begin{center}
$\Phi : (u, v) \rightarrow X^{\mu}(u, v)$ 
\end{center}
Recall that the Lagrangian of the embedding Eq. (\ref{2}) has a SU(n) symmetry such that $d = n^{2} - 1$. 

\textbf{Notation.} Let 
\begin{equation}
\begin{aligned}
&\phi^{i}(u, v) \equiv X^{\mu, i}(u, v), \hspace{0.5 cm} 
d^{d}X(u, v) = \prod_{i=1}^{n} \prod_{\mu=1}^{d} dX^{\mu, i}(u, v)  \\
&(\partial_{a}\Phi)^{\dag}(\partial_{b}\Phi) =  \sum_{i=1}^{n} (\partial_{a}\phi^{i})^{\dag}(\partial_{b}\phi^{i})  
= \sum_{i=1}^{n}  g_{\mu\nu} \partial_{a} X^{\mu, i} \partial_{b} X^{\nu, i} \\
& |\Phi|^{2} = \sum_{i=1}^{n}  g_{\mu\nu} X^{\mu, i} X^{\nu, i}, \hspace{0.5 cm}
|\Phi|^{4} = \Big(\sum_{i=1}^{n} g_{\mu\nu} X^{\mu, i} X^{\nu, i}\Big)^{2}. 
\end{aligned}
\end{equation}

\subsection{Action of abstract space}

The action should contain a part corresponding to Einstein's gravity in $\mathcal{M}$ and a
term for the embedding minimally coupled to Einstein's gravity. Let $S_{EH}$ be the Einstein Hilbert action given by
\begin{equation}
S_{EH} = \int d^{d} X(u,v) \hspace{0.05 cm} \sqrt{g} \hspace{0.05 cm} \frac{M^{2}}{2} R
\end{equation}
where $R$ is the Ricci scalar, $M$ is some mass scale 
and $\sqrt{g}$ is the determinant of the Riemannian metric $g_{\mu\nu}$.

Let $S_{\text{emb}}$ be the action of the embedding and is given by
\begin{equation}
S_{\text{emb}} = \int du \hspace{0.05 cm} dv \sqrt{-\gamma} \hspace{0.05 cm} 
 \Big[\frac{1}{2} \gamma^{ab} (\partial_{a}\Phi)^{\dag}(\partial_{b}\Phi) - V(\Phi)\Big]
\end{equation}
where $\gamma_{ab}$ is a metric on the embedded surface.
$a,b$ runs over local coordinates $u, v$ and $\sqrt{-\gamma}$ is the determinant of the metric $\gamma_{ab}$.
$S_{\text{emb}}$ is invariant under the action of an element $\text{G} \in$ SU(n).

We can also have a term non-minimally coupling embedding to gravity as
\begin{equation}
S_{\text{NM}} = \int du \hspace{0.05 cm} dv \sqrt{-\gamma} \hspace{0.05 cm} 
 \Big[\xi \Phi^{\dag}\Phi R \Big]
\end{equation}
where $\xi$ is the non-minimal coupling strength. We take $\xi = 0$ in abstract space.

Then the action of the abstract space $\mathcal{M}$ containing a two dimensional embedded surface is 
\begin{eqnarray}
\mathcal{A} &=& \int d^{d} X(u,v) \hspace{0.05 cm} \sqrt{g} \hspace{0.05 cm} 
\Big[\frac{M^{2}}{2} R + \int du \hspace{0.05 cm} dv \sqrt{-\gamma} \hspace{0.05 cm} 
 \Big(\frac{1}{2} \gamma^{ab} (\partial_{a}\Phi)^{\dag}(\partial_{b}\Phi) - V(\Phi)\Big) \Big].
\end{eqnarray}
Let
\begin{equation}
\label{14}
V(\Phi) = -\frac{\mu^{2}}{2} |\Phi|^{2} + \frac{\lambda}{4} |\Phi|^{4}.
\end{equation}

\subsubsection{Variation of the Action} Using the following formulas

\begin{equation}
\frac{\delta R}{\delta g^{\mu\nu}} = R_{\mu\nu}, \hspace{0.1 cm}
\frac{1}{\sqrt{g}} \frac{\delta \sqrt{g}}{\delta g^{\mu\nu}} = -\frac{1}{2} g_{\mu\nu}, \hspace{0.1 cm}
\frac{\delta g_{\alpha\beta}}{\delta g^{\mu\nu}} = - g_{\alpha\mu} g_{\beta\nu}
\end{equation}
and substituting $\frac{\delta \mathcal{A}}{\delta g^{\mu\nu}} = 0$, we get

\begin{equation}
\label{16}
\begin{aligned}
& \frac{M^{2}}{2} R_{\mu\nu} - \frac{1}{2} g_{\mu\nu}  \Big[\frac{M^{2}}{2} R
+ \int du \hspace{0.05 cm} dv \sqrt{-\gamma} 
\hspace{0.05 cm}  \Big(\frac{1}{2} \gamma^{ab} (\partial_{a}\Phi)^{\dag}(\partial_{b}\Phi) - V(\Phi)\Big) \Big] -
  \sum_{i=1}^{n}  \\
 &\int du \hspace{0.05 cm} dv \sqrt{-\gamma} \hspace{0.05 cm}
\Big[ \frac{1}{2}\gamma^{ab} \partial_{a} X_{\mu, i} \partial_{b} X_{\nu, i} + \frac{\mu^{2}}{2}  X_{\mu, i} X_{\nu, i} 
- \frac{\lambda}{4} (2 \hspace{0.05 cm} g_{\mu\nu} X^{\mu, i} X^{\nu, i})  (X_{\mu, i} X_{\nu, i}) \Big] = 0. 
\end{aligned}
\end{equation}

Denote
\begin{equation}
\phi^{\mu\nu, i} \equiv X^{\mu, i} X^{\nu, i} \hspace{0.5 cm} \phi_{\mu\nu, i} \equiv X_{\mu, i} X_{\nu, i}
\hspace{0.5 cm} \phi_{a\mu b\nu, i} \equiv \partial_{a} X_{\mu, i} \partial_{b} X_{\nu, i}
\end{equation}
and rewrite Eq. (\ref{16}) as
 
\begin{equation}
\begin{aligned}
& \frac{M^{2}}{2} R_{\mu\nu}  - \frac{1}{2} g_{\mu\nu} 
 \Big[\frac{M^{2}}{2} R + \int du \hspace{0.05 cm} dv \sqrt{-\gamma} \hspace{0.05 cm} 
 \Big(\frac{1}{2} \gamma^{ab} (\partial_{a}\Phi)^{\dag}(\partial_{b}\Phi) - V(\Phi)\Big) \Big] 
\\ & -   \sum_{i=1}^{n}  \int du \hspace{0.05 cm} dv \sqrt{-\gamma} \hspace{0.05 cm}
\Big[ \frac{1}{2}\gamma^{ab} \phi_{a\mu b\nu, i} + \frac{\mu^{2}}{2} \phi_{\mu\nu, i} 
- \frac{\lambda}{4} 2 \hspace{0.05 cm} g_{\mu\nu} \phi^{\mu\nu, i}  \phi_{\mu\nu, i} \Big] = 0. 
\end{aligned}
\end{equation}
This is the equation in the abstract space before the symmetry breaking.

\subsection{Action after Symmetry Breaking} The action $S_{\text{emb}}$ has SU(n) symmetry.
\label{breaking}
Let $\Phi$ acquires a VEV as
\begin{equation}
\langle\Phi\rangle = \frac{\mu}{\sqrt{\lambda}} \hspace{0.05 cm} X^{t} = v \hspace{0.05 cm} X^{t}
\end{equation}
and the SU(n) symmetry is spontaneously broken. $X^{t}$ becomes the \textit{arrow of time}.
There are certain directions $X^{\mu}$ that satisfy
\begin{equation}
\label{20}
[X^{\mu}, X^{t}] = 0.
\end{equation}
These directions form the generators of the left over symmetry group of the embedding; SU(m)$\times$U(1)
where U(1) symmetry is generated by $X^{t}$. There are $D = m^{2} - 1$ directions $X^{\mu}$ that satisfy Eq. (\ref{20}) 
and together with $X^{t}$ form a closed \textit{unitary} spacetime. Denote $X^{t} \equiv t$.
Note that after the symmetry breaking, dimensions $X^{\nu}$ that do not satisfy Eq. (\ref{20}) are
not changing with $(u,v)$. Thus, we can say that they are independent of $(u, v)$. Let $d^{\prime} = n^{2}-m^{2}-1$.
Our original abstract space $\mathcal{M}$ is now a $(1+ D + d^{\prime})$ \textit{bulk} spacetime.

Denote spacelike coordinates by $\alpha, \beta$ and timelike by $t$
\begin{equation}
\begin{aligned}
\Phi &= (\phi_{1}, \cdots, \phi_{m})^{\dag} \hspace{0.5 cm} \phi^{j} : (u, v) \rightarrow X^{\alpha, j}(u, v) \\ \nonumber
\varphi &: (u, v) \rightarrow t(u, v)
\end{aligned}
\end{equation}
such that $\Phi$ is the embedding of spacelike axes and $\varphi$ is the embedding of the
timelike axis. Recall that $\xi = 0$ in the abstract space. Now we take $\xi > 0$.

Embedding as n-tuple having a SU(n) symmetry group acquired a VEV along
$t$ axis and by doing that the SU(n) symmetry gets broken to SU(m)$\times$U(1). 
Symmetry breaking of embedding along $t$ axis does two things: (i) it changes the VEV of the
timelike embedding; $\varphi$ and (ii) non-minimally couples $\varphi$ to Einstein's gravity.
Now since, $\varphi$ is unique i.e. U(1) it corresponds to the embedding of only $t$ axis.

In abstract space, embedding is minimally coupled to Einstein's gravity.
Symmetry breaking then leads to emergence of time which causes timelike embedding to acquire an
additional non-minimal coupling to gravity. Note that $(\Delta t)^{2}  \equiv t_{0}^{2} = g_{tt} X^{t} X^{t}$, 
where $t_{0}$ denotes the first moment of time after the Big Bang. As $g_{tt}$ is part of Einstein's gravity and 
$X^{t}(u, v)$ is timelike embedding, this generates a non-minimal coupling.

We represent $g_{tt} X^{t} X^{t} \equiv \xi |\varphi|^{2} R$, where $\varphi = \varphi(u, v)$ and $|\varphi|^{2} = |X^{t}|^{2}$.  
This is our qualitative physical motivation for $\xi > 0$. Here, we assume that the non-minimal coupling parameter $\xi$ is positive.

Set $M = M_{\text{Pl}}$. The action $\mathcal{A}$ after the symmetry breaking is
\begin{equation}
\label{21}
\begin{aligned}
\mathcal{A} &= \int d^{d^{\prime}} X \hspace{0.05 cm} dt(u,v) \hspace{0.05 cm} d^{D} X(u,v)
\sqrt{-g} \hspace{0.05 cm} \Big[\frac{M_{\text{Pl}}^{2}}{2} R +  \\ & \int du dv \sqrt{-\gamma} 
 \Big(\frac{1}{2} \gamma^{ab}[ (\partial_{a}\Phi)^{\dag}(\partial_{b}\Phi) + (\partial_{a}\varphi)^{\dag}(\partial_{b}\varphi)]
  - V_{1} - V(\varphi) \Big) \Big]
\end{aligned}
\end{equation}
where 
\begin{equation}
\label{22}
\begin{aligned}
&(\partial_{a}\Phi)^{\dag}(\partial_{b}\Phi) 
=  \sum_{j=1}^{m} g_{\alpha\beta} \partial_{a} X^{\alpha, j} \partial_{b} X^{\beta, j}; \hspace{0.5 cm}
(\partial_{a}\varphi)^{\dag}(\partial_{b}\varphi) =  g_{tt} \partial_{a} X^{t} \partial_{b} X^{t} \\
&V_{1}  = -\frac{\mu^{2}}{2}  \sum_{j=1}^{m} g_{\alpha\beta} X^{\alpha, j} X^{\beta, j} 
+ \frac{\lambda}{4} \Big[ \sum_{j=1}^{m} g_{\alpha\beta} X^{\alpha, j} X^{\beta, j} \Big]^{2} \\
&V(\varphi) = \frac{\lambda}{4}( \varphi^{\dag}\varphi - v^{2})^{2} - \xi |\varphi|^{2} R
=   \frac{\lambda}{4}\big( g_{tt}X^{t}X^{t}  - v^{2}\big)^{2} 
- \xi |X^{t}|^{2} R
\end{aligned}
\end{equation}
First note that there are $D+1$ dimensions that depend on embedding; $(u, v)$. For spacelike dimensions
embedding is $m-$ tuple, thus, $j$ runs from 1 to $m$. Time dimension is uniquely embedded.
Also note that $V_{1}$ includes the potential for the spacelike directions that do not depend on $(u, v)$
and the index $j$ will be absent for these axes. They will also come outside of the second integral in Eq. (\ref{21}). 

\subsubsection{Variation of the Action}
Next we vary the action $\delta \mathcal{A}$ and get
\begin{equation}
\label{23}
\begin{aligned}
& \frac{M_{\text{Pl}}^{2}}{2}R_{\mu\nu} -\frac{1}{2} g_{\mu\nu}  
\Big[\frac{M_{\text{Pl}}^{2}}{2} R +  \int dudv \sqrt{-\gamma} \\
& \Big(\frac{1}{2} \gamma^{ab} [(\partial_{a}\Phi)^{\dag}(\partial_{b}\Phi) + (\partial_{a}\varphi)^{\dag}(\partial_{b}\varphi)] 
- V_{1} - V(\varphi) \Big) \Big]  \\
& - \int du dv \sqrt{-\gamma}  \frac{1}{2}\gamma^{ab} \Big[\partial_{a} X_{\mu} \partial_{b} X_{\nu}\Big] 
 -  \int du dv \sqrt{-\gamma} \\
& \Big[ \Big(\frac{\mu^{2}}{2} (1 - \delta_{t\mu} \delta_{t\nu}) - \frac{2 \lambda}{4} \big[g_{\mu\nu} X^{\mu} X^{\nu}
- v^{2} \delta_{t\mu} \delta_{t\nu} \big] \Big) X_{\mu} X_{\nu}  
  - \xi |X^{t}|^{2} R_{\mu\nu} \Big] = T_{\mu\nu} 
\end{aligned}
\end{equation}
where $\delta_{ij}$ is the delta function. Spacelike dimensions that depend upon embedding carry an index $j$ as well.
$T_{\mu\nu}$ is the stress-energy tensor which is non zero since there is at least a vacuum like
energy, $\Delta E$ present in the universe, due to the uncertainty principle,
\begin{equation}
\begin{aligned}
\Delta E \Delta t &\geqslant \hbar.
\end{aligned}
\end{equation}

We can rewrite Eq. (\ref{23}) as
\begin{equation}
\begin{aligned}
&\frac{M_{\text{Pl}}^{2}}{2} R_{\mu\nu} -\frac{1}{2} g_{\mu\nu} 
\Big[\frac{M_{\text{Pl}}^{2}}{2} R + \int du \hspace{0.05 cm} dv \sqrt{-\gamma} \\
& \Big(\frac{1}{2} \gamma^{ab} [(\partial_{a}\Phi)^{\dag}(\partial_{b}\Phi)  + (\partial_{a}\varphi)^{\dag}(\partial_{b}\varphi)] 
  - V_{1} - V(\varphi) \Big) \Big]  \\
& - \int du dv \sqrt{-\gamma} \frac{1}{2}\gamma^{ab} 
 \Big[\phi_{a\mu b\nu} \Big]  -  \int du dv \sqrt{-\gamma}  \\
& \Big[\Big(\frac{\mu^{2}}{2}(1 - \delta_{t\mu} \delta_{t\nu}) - \frac{2  \lambda}{4}
\big[g_{\mu\nu}\phi^{\mu\nu}  - v^{2} \delta_{t\mu} \delta_{t\nu} \big] \Big) \phi_{\mu\nu} 
 - \xi |X^{t}|^{2} R_{\mu\nu}  \Big] = T_{\mu\nu}. 
\end{aligned}
\end{equation}

\section{Discussion}
\label{discussion}
There are somewhat subtle and important manifestations of the new physics associated with embedding.
They are qualitatively explained below. 

\subsection{Expansion and unitary evolution}
\label{expansion}
Note that after the symmetry breaking the directions $X^{\nu}$ such that $[X^{\nu}, t] \neq 0$ 
are non dynamical. They do not depend on the coordinates $(u, v)$ of the embedded surface. 
Embedding provides a mechanism for the dimensions to expand or contract depending upon the
energy density as predicted by the Einstein's equations. To better understand this mechanism
let $\sigma \equiv (u, v)$. For any dimension say, $X^{\alpha}(\sigma)$, the differential
$\frac{\delta X^{\alpha}}{\delta \sigma}$ is in general $\neq 0$. This gives a dynamical meaning to $X^{\alpha}(\sigma)$.
Since, $\frac{\delta t}{\delta \sigma} \neq 0$ we can in general have $\frac{\delta X^{\alpha}}{\delta t} \neq 0$.
Whereas for $X^{\nu}$ independent of $\sigma$ we have $\frac{\delta X^{\nu}}{\delta \sigma} = 0$ always
and thus, $\frac{\delta X^{\nu}}{\delta t} = 0$ always.
Hence, the dimensions $X^{\nu}$ remain finite 
(assuming they start out as finite). Therefore, when Einstein wrote down his equations
there was a-priori no reason for him to expect the spacetime to be dynamical. As his equations
are dynamical and the observations confirm an expanding universe it is the embedding that 
causes the three space dimensions to expand creating more space. 

Let $X^{\mu}$ such that $[X^{\mu}, t] = 0$ denote the space like dimensions that 
are dynamical, i.e. depend on embedding. Since they commute with time the dynamics 
in the spacetime along these axes is transparent to the time axis. This transparency manifests itself 
as \textit{unitary evolution} in our spacetime. To understand what it means for dynamics to be transparent let 
$X^{1}, X^{2} \in \{X^{\mu}\}$. Then,
\begin{eqnarray}
&&[t, X^{1}X^{2}] = [t, X^{1}] X^{2} + X^{1} [t, X^{2}] = 0 \nonumber \\
&&[t, X^{2}X^{1}] = [t, X^{2}] X^{1} + X^{2} [t, X^{1}] = 0.
\end{eqnarray}
This is the unitary evolution in the subspace formed by $\{t, X^{\mu}\}$.

\subsection{Example : SU(3) case}

As an example, consider $d = 8$; a $(0, 8)$ abstract manifold $\mathcal{M}$ (\textit{bulk} space).
The embedding of the dark surface is $\Phi = \{\phi^{1}, \phi^{2}, \phi^{3}\}^{\dag}$, with
\begin{equation}
\phi^{i} : (u, v) \rightarrow (X^{1, i}(u, v), X^{2, i}(u, v), \cdots, X^{8, i}(u, v)). \nonumber
\end{equation}
where $i = 1, 2, 3$. For $\Phi \rightarrow \mathrm{G} \Phi$, where $\mathrm{G} \in$ SU($3$), 
$S_{\text{emb}}$ is invariant. Then $X^{\mu}$ are the generators of SU($3$); Gell-mann matrices;
\begin{flushleft}
$ X^{1} = \left(\begin{array}{ccc}0 & 1 & 0 \\1 & 0 & 0 \\0 & 0 & 0\end{array}\right)$,
$ X^{2} = \left(\begin{array}{ccc}0 & -i & 0 \\i & 0 & 0 \\0 & 0 & 0\end{array}\right)$,
$ X^{3} = \left(\begin{array}{ccc}1 & 0 & 0 \\0 & -1 & 0 \\0 & 0 & 0\end{array}\right)$,
$ X^{4} = \left(\begin{array}{ccc}0 & 0 & 1 \\0 & 0 & 0 \\1 & 0 & 0\end{array}\right)$,
$ X^{5} = \left(\begin{array}{ccc}0 & 0 & -i \\0 & 0 & 0 \\i & 0 & 0\end{array}\right)$,
$ X^{6} = \left(\begin{array}{ccc}0 & 0 & 0 \\0 & 0 & 1 \\0 & 1 & 0\end{array}\right)$,
$ X^{7} = \left(\begin{array}{ccc}0 & 0 & 0 \\0 & 0 & -i \\0 & i & 0\end{array}\right)$,
$ X^{8} = \frac{1}{\sqrt{3}}\left(\begin{array}{ccc}1 & 0 & 0 \\0 & 1 & 0 \\0 & 0 & -2\end{array}\right)$.
\end{flushleft}
Let $\Phi$ acquires a VEV as
\begin{equation}
\label{27}
\langle \Phi \rangle =  \frac{\mu}{\sqrt{\lambda}} \left(\begin{array}{ccc}1 & 0 & 0 \\0 & 1 & 0 \\0 & 0 & -2\end{array}\right).
\end{equation}

As this direction commutes with $X^{1}, X^{2}, X^{3}$ and $X^{8}$ the SU(3) symmetry is spontaneously broken 
to SU(2)$\times$U(1). The \textit{bulk} space $\mathcal{M}$ is now a (1+ 3 + 4) dimensional \textit{bulk} spacetime.
A frame chosen along $\{X^{8}, X^{1}, X^{2}, X^{3}\}$
preserves the unitary dynamics. They depend on embedding and form a closed set of non orthogonal basis. 
Whereas dimensions $X^{4}, X^{5}, X^{6}$ and $X^{7}$ do not depend on embedding and hence, are non dynamical
since $\frac{\delta X^{\nu}}{\delta t} = 0$ where $\nu = 4, 5, 6, 7$ as discussed in \ref{expansion}. 
Inflation has no effect on them. They remain finite even when the universe undergoes exponential expansion.
Due to the closed structure of $\{X^{8}, X^{1}, X^{2}, X^{3}\}$
and their dependence on the embedding, classically i.e. after inflation they seem to be complete. 
Identifying, $X^{8}$ as $t$ axis, and $X^{1}, X^{2}, X^{3}$ as $x, y, z$ respectively we get a classical spacetime $\{t, x, y, z\}$.
Thus, our observable universe is 1+ 3 dimensional.

\subsubsection{Inflationary scenario}
\label{inflation}
Embedding gravity is introduced here as new physics that modifies gravity at Planck scale. 
Initially, embedding has 3 flavors and SU(3) symmetry with VEV as zero.
Embedding then acquires a non-zero component along $t$ axis, during spontaneous symmetry breaking;
SU(3)$\rightarrow$SU(2)$\times$U(1). The VEV now is non-zero Eq. (\ref{27}).
Note that since the displacement of embedding is along the time axis, the timelike embedding has a non-zero VEV.
In addition, as explained earlier in \ref{breaking} it also acquires a non-minimal coupling to Einstein's gravity with
the emergence of time. Due to these two things timelike embedding experiences a different potential. 
This different potential then causes timelike embedding (or embedding of $t$ axis, since it is unique i.e. only 1 flavor) 
to drive inflation. 

Denote
\begin{equation}
\begin{aligned}
&\Phi = (\phi_{1}, \phi_{2})^{\dag} \\ 
&\phi^{i} : (u, v) \rightarrow (X^{1, i}(u, v), X^{2, i}(u, v), X^{3, i}(u, v)) \\
&\varphi :  (u, v) \rightarrow X^{8} (u, v)     \nonumber
\end{aligned}
\end{equation}
Let $v = \frac{\mu}{\sqrt{\lambda}}$ and from Eq. (\ref{22}) with $1 \leqslant \alpha, \beta \leqslant 3$ (unitary subspace)
\begin{equation}
\begin{aligned}
V(\Phi) &= -\frac{\mu^{2}}{2} |\Phi|^{2} + \frac{\lambda}{4} |\Phi|^{4}
= -\frac{\mu^{2}}{2}  \sum_{i=1}^{2} g_{\alpha\beta} X^{\alpha, i} X^{\beta, i} 
+ \frac{\lambda}{4} \Big[ \sum_{i=1}^{2} g_{\alpha\beta} X^{\alpha, i} X^{\beta, i} \Big]^{2}   \\ \nonumber
V(\varphi) &= \frac{\lambda}{4} (\varphi^{\dag}\varphi - v^{2})^{2} - \xi |\varphi|^{2} R
= \frac{\lambda}{4}\big( g_{88}X^{8}X^{8}  - v^{2}\big)^{2} - \xi |X^{8}|^{2} R
\end{aligned}
\end{equation}
where $\xi$ is the non-minimal coupling strength. (See \cite{BS} for the role of non-minimal coupling in Higgs inflation). 

For $\varphi^{\dag}\varphi \gg v^{2}$ i.e. at large field values $V(\varphi)$ becomes
\begin{equation}
\label{28}
V_{1}(\varphi) =  \frac{\lambda}{4} (\varphi^{\dag}\varphi)^{2} - \xi |\varphi|^{2} R.
\end{equation}
Then the action of the unitary subspace (that undergoes inflation) of the bulk spacetime $\mathcal{M}$ is
\begin{equation}
\label{29}
\begin{aligned}
\mathcal{A} = \int dt(u,v) \hspace{0.05 cm} d^{3} X(u,v)
\sqrt{-g} \hspace{0.05 cm} \Big[\frac{M_{\text{Pl}}^{2}}{2} R +  &
 \int du dv \sqrt{-\gamma}
 \Big(\frac{1}{2} \gamma^{ab} (\partial_{a}\Phi)^{\dag}(\partial_{b}\Phi)  - V (\Phi) \\
& + \frac{1}{2} \gamma^{ab}(\partial_{a}\varphi)^{\dag}(\partial_{b}\varphi) - V_{1}(\varphi) \Big) \Big] 
\end{aligned}
\end{equation}

Note that a conformal rescaling to Einstein's frame as $\tilde{g}_{\mu\nu} = \Omega^{2} g_{\mu\nu}$, means
$\Omega^{2}$ is independent of the metric $g_{\mu\nu}$. Thus, at the emergence of time (i.e. after symmetry breaking)
this constraint on $\Omega^{2}$ also forces $g_{88} X^{8} X^{8}$ to act as a non-minimal coupling. 
Under conformal transformation
\begin{equation}
\tilde{g}_{\mu\nu} = \Omega^{2} g_{\mu\nu}, \hspace{0.5 cm} 
\Omega^{2} \equiv 1 +  \int du dv \sqrt{-\gamma} \hspace{0.05 cm} \frac{2\hspace{0.05 cm} \xi |\varphi|^{2}}{M_{\text{Pl}}^{2}} 
\end{equation}
and for $\xi \gg 1$, 
the potential in Eq. (\ref{28}) is known to be identical to Starobinsky $R^{2}$ model (see for instance \cite{BM})
\begin{equation}
V_{1}(\varphi) = \frac{\lambda M_{\text{Pl}}^{4}}{4 \xi^{2}} \Big(1 - \text{exp}\hspace{0.05 cm} 
\Big[- \sqrt{\frac{2}{3}}\frac{\varphi}{M_{\text{Pl}}} \Big] \Big)^{2}.
\end{equation}
Then the model with constraint $\xi = 47000 \sqrt{\lambda}$ has the same phenomenology
as that of Starobinsky $R^{2}$ model of inflation with associated slow roll parameters as 
\cite{BM}
\begin{equation}
\eta = - \frac{4}{3} e^{-\sqrt{2/3}\varphi/M_{\text{Pl}}}, \hspace{0.5 cm}
\epsilon = \frac{3}{4} \eta^{2}.
\end{equation}
The observables spectral tilt and
tensor to scalar ratio are also same as that of Starobinsky $R^{2}$ model \cite{Star}
\begin{equation}
\label{33}
n_{s} \approx 1 - \frac{2}{N}, \hspace{0.5 cm} r \approx \frac{12}{N^{2}} 
\end{equation}
where $N$ is the number of e-foldings since the horizon crossing. For $50 < N < 60$, the observables Eq. (\ref{33})
are in excellent agreement with the Planck data \cite{Planck}. 

The inflationary scenario discussed here with (timelike) embedding as inflaton is
a feature of gravity at Planck scale. It is the embedding gravity that drives inflation instead of some
arbitrary scalar field. While our picture is conceptually closer to that of 
Starobinsky $R^{2}$ inflation \cite{Star} in a sense it modifies gravity at Planck scale, 
embedding's slow-roll mechanism to drive the inflation was first proposed in the models \cite{L1, AS}.
Also note that other scenarios such as Higgs inflation, universal attractor models that agree with the Planck data
\cite{CMRV} are shown to be Starobinsky $R^{2}$ model during inflation \cite{KDR, HSY}.

\subsection{Effect of Inflation}

As discussed in \ref{inflation} after the symmetry breaking inflation naturally starts in the unitary subspace $\{t, x, y, z\}$
of the bulk spacetime $\mathcal{M}$. Note that
\begin{equation}
[x, y] = 2 \hspace{0.05 cm} i \hspace{0.05 cm} z
\end{equation}
and from Heinsenberg's minimum uncertainty principle, 
\begin{equation}
\Delta x \Delta y = \frac{1}{2} \langle [x, y] \rangle.
\end{equation}
Thus, $\langle [x, y] \rangle = 2 \hspace{0.05 cm} i \hspace{0.05 cm} \langle z \rangle$.
Let $ \langle x \rangle =  \langle y \rangle =  \langle z \rangle = \theta_{0}$ be constant. Then
\begin{equation}
\Delta x \Delta y = \theta_{0}.
\end{equation}
Following \cite{CGS} in comoving coordinates, 
\begin{equation}
\theta(t) = \frac{\theta_{0}}{a(t)}
\end{equation}
that is the uncertainty in position measurements (hence, the non-commutativity) gets redshifted away with inflation.
To study the non-commutativity in physical coordinates, let the distance between any two points be $\Lambda$.
It has an uncertainty of $\mathcal{O}(\sqrt{\theta_{0}})$. With expansion,
\begin{equation} 
\Lambda (t) = a(t) \Lambda.
\end{equation}
Due to inflation, $\Lambda (t) \gg \sqrt{\theta_{0}}$ as $\theta_{0}$ is constant and the non-commutativity dies away.
Thus, for classical spacetime to be commutative the non-commutativity parameter $\theta_{0}$ has to be a constant.
As the non-commutativity parameter $\theta_{0}$ depends on the uncertainty in position measurements,
this means that the uncertainty in the position measurements is also constant. But for this uncertainty to be constant 
the universe must be \textit{homogeneous}. If for example, $\langle z  \rangle = f(x,y, z)$,
then the space is inhomogeneous, the parameter $\theta_{0}$ is not constant and non-commutativity stays after the inflation. 
This is because the uncertainty in position measurements (and hence $\theta_{0}$) 
also scales as \textit{a(t)} and therefore, $\Lambda (t) \sim \sqrt{\theta_{0}}$ after inflation.

\section{Conclusion}
\label{conclusion}
Embedding of a surface (this surface is not a brane on which our visible universe is restricted)
in an abstract \textit{bulk} space when studied as a map under the Sigma model framework
(unlike the braneworld cosmology that characterizes embedding by extrinsic curvature)
has a rich physics. 
The new physics associated with embedding cosmology provides a physical explanation 
for the Big Bang, the arrow of time and the dark sector. The embedded dark surface may correspond to what is known as
the dark matter in the \textit{bulk} whereas the energy in the vacuum state of embedding may offer an explanation for 
the dark energy.

Embedding gravity introduced here as new physics at Planck scale is shown to drive inflation. 
Symmetry breaking causes timelike embedding to acquire an additional
non-minimal coupling to gravity with coupling parameter $\xi$ assumed to be positive. Thus, it is the embedding of
the time axis that acts as inflaton instead of some arbitrary scalar field. 
It is shown to have same phenomenology as Starobinsky $R^{2}$ inflation. 
Embedding also explains why the extra dimensions near Planck scale maybe finite as they lack the
machinery required for those space dimensions to expand and therefore, explains why the observable
universe is 1+3 dimensional.

\textbf{Acknowledgement}. I would like to thank Will Kinney and Dejan Stojkovic for helpful discussions
during the early stages of the work. 
I would also like to thank Jonathan Dimock for advising my Ph.D dissertation.

\end{document}